\documentclass[twocolumn,showpacs,preprintnumbers,amsmath,amssymb,prb,aps]{revtex4-1}
\usepackage{epsfig}
\usepackage{graphicx}
\usepackage{dcolumn}
\usepackage{bm}
\usepackage{textcomp}
\usepackage{subfig}

\begin{document}

\title{Investigation of Thermoelectric Properties of ZnV$_{2}$O$_{4}$ Compound at High Temperatures}

\author{Saurabh Singh${{^{1}}}$}
\altaffiliation{Electronic mail: saurabhsingh950@gmail.com}
\author{R. K. Maurya${{^{2}}}$}
\author{Sudhir K. Pandey${{^{1}}}$}
\affiliation{${{^{1}}}$School of Engineering, Indian Institute of
Technology Mandi, Kamand - 175005, India}

\affiliation{${{^{2}}}$School of Basic Sciences, Indian Institute of
Technology Mandi, Kamand - 175005, India}

\date{\today}

\begin{abstract}
In the present work, we report the experimental thermopower ($\alpha$) data for ZnV$_{2}$O$_{4}$ in the high temperature range 300-600 K. The values of $\alpha$ are found to be $\sim$184 and $\sim$126 $\mu$V/K at $\sim$300 and $\sim$600 K, respectively. The temperature dependent behavior of $\alpha$ is almost linear in the measured temperature range. In order to understand the large and positive $\alpha$ values observed in this compound, we have also investigated the electronic and thermoelectric properties by combining the \textit{ab-initio} electronic structures calculations with Boltzmann transport theory. Within the local spin density approximation plus Hubbard U, the anti-ferromagnetic ground state calculation gives an energy gap $\sim$0.33 eV for U=3.7 eV, which is in accordance with the experimental results. The effective mass for holes in the valence band is found nearly four times that of electrons in conduction band. The large effective mass of holes are mainly responsible for the observed positive and large $\alpha$ value in this compound. There is reasonably good matching between calculated and experimental $\alpha$ value in the temperature range 300-410 K. The power factor calculation shows that thermoelectric properties in high temperature region can be enhanced by tuning the sample synthesis conditions and suitable doping. The estimated value of \textit{figure-of-merit}, \textit{ZT}, for \textit{p-type} doped ZnV$_{2}$O$_{4}$ is $\sim$0.3 in the temperature range 900-1400 K. It suggests that by appropriate amount of \textit{p-type} doping, this compound can be a good thermoelectric material in high temperature region.

\end{abstract}


\maketitle

\section{Introduction} 
In the past few decades, research is being focused on developing the functional materials which can utilize the various energy sources such as solar, wind, hydro-power, biomass and thermal energy for electrical energy production.\cite{DiSalvo,Bell} The various technologies used to generate electricity from these sources produce large amount of waste heat in the environment. Therefore, utilization of thermal energy for electricity production have attracted much attention as they directly convert thermal energy into electrical energy and vice versa.\cite{Snyder,Trit,Rowe} Thermoelectric energy conversion devices made by using the thermoelectric materials have various applications such as power generation, temperature sensors, thermoelectric cooler for electronic devices, cooling infrared sensors, etc.\cite{Riffat} A thermoelectric device is efficient for thermoelectric applications decided by the dimensionless parameter, known as \textit{figure-of-merit} (\textit{ZT}). The value of \textit{ZT} depends on transport properties of material and defined as,\cite{Pei,LaLonde}
\begin{equation}
  \textit{ZT}= \alpha^{2}\sigma T/\kappa 
\end{equation}
 where $\alpha$, $\sigma$, $\kappa$, and T are Seebeck coefficient (also known as thermopower), electrical conductivity, thermal conductivity ($\kappa_{e}$ + $\kappa_{l}$), and absolute temperature, respectively. The symbol $\kappa_{e}$ and $\kappa_{l}$ represents the electronic and lattice thermal conductivity, respectively. To make an efficient thermoelectric energy conversion device, the materials should have high $\sigma$, large $\alpha$, and low $\kappa$.\cite{Nolas} The low $\kappa$ is essential to establish a large temperature gradient between two ends of the material, whereas large value of $\alpha$ is required to generate a high voltage per unit temperature gradient and high $\sigma$ is needed to reduce the internal resistance of the material.\\ 
 In the search of good thermoelectric materials with consideration of above criteria, the various thermoelectric materials have been explored which are being used to make the thermoelectric devices.\cite{Snyder} The state-of-the-art thermoelectric materials used in thermoelectric applications are Be-Te-based alloys (\textit{ZT}$\approx$ 0.8 to 1.4) for both \textit{n} and \textit{p}-type thermoelectric systems, they are mostly useful for refrigeration and waste heat recovery upto 450 K temperature.\cite{Poudel} For the intermediate temperature region, i.e., 500 K to 900 K, Pb-Te-based alloys ( maximum \textit{ZT} $>$ 1.5 at 800 K) are more suitable.\cite{Heremans} For higher temperature ($>$900 K), silicon and germanium based alloys are used in making the thermoelectric generators. Although Be-Te, Pb-Te based alloys and silicon-germanium alloys remains the high demanding materials for making the commercial thermoelectric generator and refrigerator, but the environmental issues of hazardous Bi and Pb alloys restricts them from several applications. Further, these materials are not suitable for high temperatures due to their easy decomposition and oxidation in the air. Moreover, use of these materials should be avoided as far as possible, as they are toxic, and not environment friendly. In Comparison to these thermoelectric alloys, thermoelectric oxides are more suitable for high-temperature applications because of their low manufacturing cost, environmental friendliness, nontoxic character, availability in nature, structural and chemical stabilities, oxidation resistance.\cite{Koumoto, KKoumoto}\\
 In 1997, the discovery of large and positive thermopower in Na$_{x}$CoO$_{2}$ compound gave a new breakthrough for oxides materials as potential thermoelectric candidates for high temperature range applications.\cite{Terasaki} In the last few decades transition-metal oxides with strongly correlated electron systems have attracted much attention due to spin, charge, orbital and lattice degree of freedom of the electrons.\cite{Jefferey,Swalia, Ohtaki} Among the discovered transition-metal oxides, the small-band gap semiconductor oxides are suitable for thermoelectric applications as they have large thermopower, and electrical conductivity. ZnV$_{2}$O$_{4}$ belonging to the spinel oxides have a small energy band gap and its experimentally reported value is $\sim$0.32 eV.\cite{Rogers} The electronic structure of this compound has been investigated by the various group. They have studied the possibility of orbital ordering responsible for the structural transition from cubic to tetragonal observed in this compound.\cite{Tchernyshyov, Khomskii, Maitra, Lal, Motome, Matteo, Lee} For the similar single crystal spinel oxides, the value of thermal conductivites are reported by Ishitsuka et al., and $\kappa$ values at $\sim$200 K are $\sim$3.33 W/mK for CoV$_{2}$O$_{4}$ and MnV$_{2}$O$_{4}$ compound.\cite{Ishitsuka} ZnV$_{2}$O$_{4}$ shows large and positive thermopower $\sim$350 $\mu$V/K at $\sim$200 K.\cite{Canosa} The value of $\alpha$ decreases almost linearly with increase in temperature. As per best of our knowledge, thermoelectric properties of this compound has not been studied above 400 K. The detailed analysis of the various physical parameters which are responsible for the positive and large thermopower shown by this compound is still lacking in the literature. Thus, ZnV$_{2}$O$_{4}$ can be thermally stable and a good thermoeletric material in higher temperature range. This gives motivation to explore thermoelectric properties of this compound using theoretical and experimental tools.\\
 In this paper, we have studied the thermopower behaviour of ZnV$_{2}$O$_{4}$ in the temperature range 300-600 K with the help of experimental tools. The electronic properties has been studied using the first principle density functional theory. The total density of states (TDOS) has been calculated for the anti-ferromagnetic tetragonal unit cell. In order to see the contributions of charge carriers from different atomic bands to thermopower, we have also plotted the partial density of states (PDOS) of V $\textit{3d}$ and O \textit{2p} orbitals. In order to understand the large and positive thermopower exhibited by this compound we have estimated the effective masses of holes and electrons. Thermopower data is also obtained using BoltzTraP calculation in the temperature range 200-1400 K. The power factor (PF) and \textit{figure-of-merit}, \textit{ZT}, are estimated theoretically to see the potential capability of the compound to use it as a thermoelectric material in the appropriate region of high temperature range.

  \section{Experimental and Computational details }
  
  Polycrystalline ZnV$_{2}$O$_{4}$ sample was prepared by using the conventional solid state ceramic route.\cite{Reehuis} The starting materials, ZnO($\geq$ 99.999\% purity) and V$_{2}$O$_{3}$($\geq$ 99.99\% purity) were taken in the stoichiometric ratio. To get the homogeneous mixture, starting materials were grounded for 6 hour by using the mortar and pestle. Further, using the homogeneous mixture, pellet of 10 mm diameter was made under the pressure of $\sim$120 Kg/cm$^{2}$. Then, the pellet was sealed in quartz tube and sintered at 800 $^{0}$C for 24 hours under the vacuum condition ($\sim$10$^{-6}$ mbar). The experimental thermopower data on ZnV$_{2}$O$_{4}$ was obtained in the temperature range 300-600 K by using the home-made thermopower measurement setup.\cite{Singh} For thermopower measurement, sample was used in the pellet form having the thickness of $\sim$1 mm and diameter of $\sim$10 mm.\\
  The spin-polarized LSDA+U calculations have been carried out employing the state-of-the art full-potential linearized augmented plane-wave method using WIEN2K software.\cite{Blaha}  LSDA+U approximation have been used to account the electronic correlations, where U is the on-site Coulomb interaction strength among V \textit{3d} electrons.\cite{Perdew} The BoltzTrap code is used to calculate the temperature dependent transport coefficients within the constant relaxation time approximations, which has been used with notable success to describe the thermopower of a large number of thermoelectric materials.\cite{Madsen} In the calculation, we have used the experimental lattice parameters a=b= 5.9526 $\AA$, c= 8.3744 $\AA$ of tetragonal phase which is corresponding to I41/amd space group.\cite{Reehuis} At this point it is important to notice that, Pardo et al. have reported the dimerize structure of ZnV$_{2}$O$_{4}$ for small value of U$_{eff}$ (below about 3 eV).\cite{Pardo, Baldomir} They have suggested that, dimerized structure corresponding to P4$_{1}$2$_{1}$2 (No. 92) space group is more stable than the standard reported structure. However, the said structure has not been reported experimentally so far to the best of our knowledge. In the literature, Reehuis et al. have reported that structure of this compound is tetragonal with space group I41/amd, and this is widely accepted structure in the research community.\cite{Reehuis} Therefore, for ZnV$_{2}$O$_{4}$ we have chosen the tetragonal structure with U value equal to 3.7 so that it gives the value of energy band gap equal to the experimentally reported band gap.\cite{Rogers} The muffin-tin sphere radii used in the calculations were set automatically to 1.99, 1.98, and 1.71 Bohr for Zn, V, and O atoms, respectively. The value of R$_{MT}$K$_{max}$, which determines the matrix size for convergence was set equal to 7, where R$_{MT}$ is the smallest atomic sphere radii and K$_{max}$ is the plane wave cut-off. The self-consistency was achieved by demanding the convergence of the total charge/cell to be less than 10$^{-4}$ electronic charge. The k-point mesh was set to a size of 30$\times$30$\times$30 during the calculations of electronic and transport properties. The value of lpfac parameter, which represents the number of k-points per lattice point, was set to 5 for calculation of thermopower. For the calculation of transport property, the value of chemical potential was taken same as obtained in the self-consistent calculations. The value of chemical potential used in the calculations of transport coefficients was 0.49036 Ry.
  
\section{Results and Discussion}
At room temperature,  ZnV$_{2}$O$_{4}$ exhibit cubic spinel structure characterized by space group \textit{Fd$\bar3$m}. A structural phase transitions from cubic to tetragonal structure takes place at T$_{s}$ = 51 K.\cite{Ueda} ZnV$_{2}$O$_{4}$ has anti-ferromagnetic structure in the ground state.\cite{Reehuis} At this stage, it is important to note that at higher temperature the compound is reported to show the paramagnetic behavior. In the paramagnetic phase of strongly correlated systems normally local magnetic moments are present. These magnetic moments are randomly oriented. In order to get the correct band structure of the compound it is necessary to perform spin-polarized calculations which will give rise to the magnetic moment at the V sites, as spin unpolarized does not consider spin dependent potential. Moreover, spin unpolarized DFT +U calculations normally unable to provide the insulating gap, which is necessary to understand the thermoelectric behavior of this compound. Here, we have performed the electronic structure calculations corresponding to tetragonal phase as it will be suitable to study temperature dependent thermoelectric behavior in wide temperature range. In present case tetragonal distortion is small (c/a = 0.9948) and hence it is expected that the splitting energy of the degenerate orbitals in the cubic phase will be in the same energy range in the higher temperature region.\cite{Slal} In this temperature range those orbitals will behave as degenerate orbitals in the cubic phase. Therefore, in the present work, we have studied electronic and thermoelctric properties in the tetragonal phase. The symmetry in tetragonal phase is described by space group \textit{I41/amd}. In this structure, the Wyckoff positions occupied by Zn, V, and O atoms are 4a (0, 3/4, 1/8), 8d (0, 0, 1/2) and 16h (0, x, z), respectively. The value of x and z, for the Oxygen atomic positions are 0.02 and 0.2611, respectively. 
 \vspace{0.1cm}
\begin{figure}[htbp]
  \begin{center}
    \includegraphics[width=0.45\textwidth]{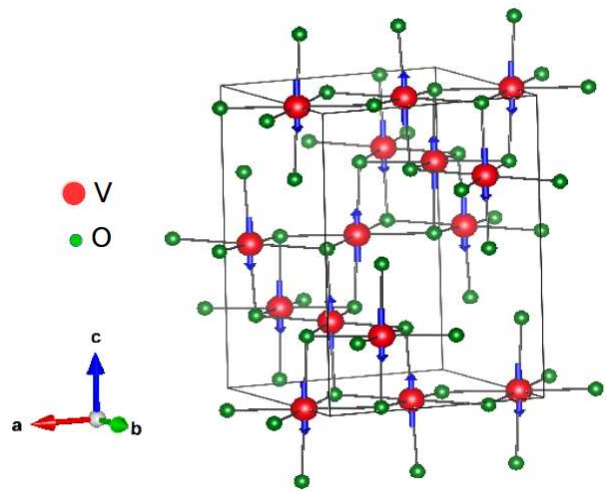}
    \label{}
    \captionsetup{justification=raggedright,
singlelinecheck=false
}
    \caption{(Color online) Antiferromagnetic structure of ZnV$_{2}$O$_{4}$ compound shown in the unit cell. The spin arrangements of V atoms are in antiferromagnetic chains ($\uparrow$$\downarrow$$\uparrow$$\downarrow$) along the \textit{\textbf{a}} and \textit{\textbf{b}}-axes. Each V atoms form the Octahedron by six surrounding O atoms. Zn atoms and O-O bonds are not shown here for the sake of clarity.}
    \vspace{-0.1cm}
  \end{center}
\end{figure} 
The tetragonal unit cell in which V$^{3+}$ ions forming the antiferromagnetic structure is shown in the Fig. 1. V$^{3+}$ (S=1) ions form a corner-sharing tetrahedral network and each V atoms are octahedrally coordinated by six oxygen ions. The spin sequences (spin up$\uparrow$ and spin dn $\downarrow$) of V$^{3+}$ ions   are arranged in such a way that it makes an antiferromagnetic chains ($\uparrow$$\downarrow$$\uparrow$$\downarrow$) along \textit{\textbf{a}} and \textit{\textbf{b}}-axes. \\
\begin{figure}[htbp]
  \begin{center}
   \includegraphics[width=0.35\textwidth, totalheight=0.45\textheight]{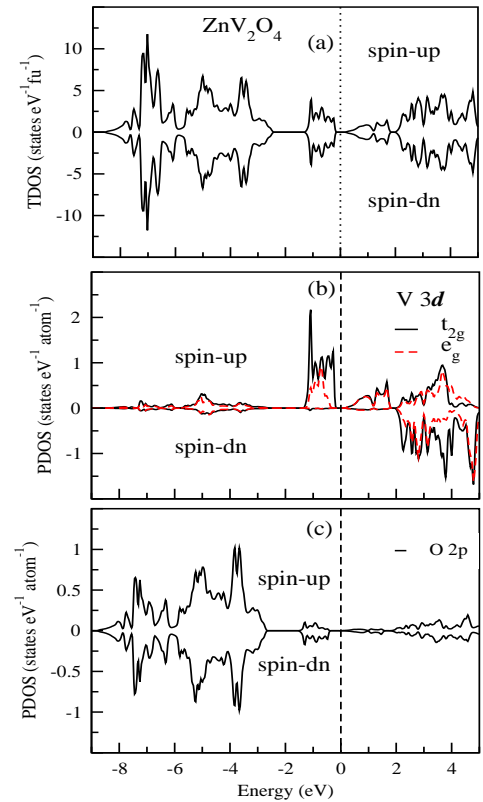}
    \label{}
    \captionsetup{justification=raggedright,
singlelinecheck=false
}
    \caption{(Color online) Total and partial density of states plots for ZnV$_{2}$O$_{4}$. Shown are (a) the TDOS plot, (b) PDOS of V atom, (c) PDOS of O atom.}
    \vspace{-0.2cm}
  \end{center}
\end{figure} 
In order to know the ground state of the system, self-consistent field calculation on this compound has been done for both ferromagnetic and anti-ferromagnetic structure. The total converged energy for anti-ferromagnetic solution is found to be $\sim$985 meV/f.u. lower than that of ferromagnetic solution. Thus, the electronic density of states have been calculated for the anti-ferromagnetic structure. The total density of states (TDOS) plot obtained in anti-ferromagnetic phase for ZnV$_{2}$O$_{4}$ is shown in Fig. 2(a). From the TDOS plot, the estimated value of energy gap is $\sim$0.33 eV. For the intrinsic semiconductor materials having the energy gap E$_{g}$, the chemical potential ($\mu$) is defined by the following formula,\cite{Ashcroft}

\begin{equation}
  \mu = \varepsilon_{v} + \frac{1}{2}Eg + \frac{3}{4}\textit{k}_{B}T \textit{ln} ({\textit{m}_{\textit{v}}}/{\textit{m}_{\textit{c}}})
  \end{equation}
  where $\varepsilon$$_{v}$ is the energy of the electron at the top of the valence band (VB), ${m_{v}}$ and ${m_{c}}$ are the effective mass of the charge carrier in valence and conduction band (CB), respectively. The above formula implies that as T $\rightarrow$ 0, the chemical potential $\mu$ lies precisely in the middle of the energy gap. In the present case, the electronic properties are calculated from the ground state self consistent solution which corresponds to the T=0 K. Therefore, we set the chemical potential in the middle of energy gap (i.e. energy difference between top of the VB and bottom of the CB). The dashed line represents the chemical potential ($\mu$), which is set at the middle of energy gap.\\
    From the TDOS plot, we have observed an equal and opposite contributions in density of states for both up and down-spins, as it was expected for AFM phase. Thus, there is no net spin magnetic moment obtained per unit cell. For the value of U (on-site Coulomb interaction strength) equal to 3.7 eV, the estimated energy gap is found to be 0.33 eV which is same as experimentally observed band gap.\cite{Rogers} In the anti-ferromagnetic structure, two V atoms distinguished by their net spin (up and dn) are present in per formula unit. The partial density of states (PDOS) plots of the V atom (\textit{3d} orbitals), having net spin up, and O atom (\textit{p} orbital) are shown in the Fig. 2(b) $\&$ 2(c), respectively. From the PDOS of V atom in Fig. 2(b), it is clear that the spin-up channel have main contribution in VB whereas spin-dn channel have negligible contribution in the density of states. The equal and opposite contributions in PDOS will be from the V atoms having the net spin down. For the 3d orbitals of V atom, three fold degenerate state (d$_{xy}$, d$_{yz}$, d$_{zx}$) and two fold degenerate state (d$_{x^{2}-y^{2}}$, d$_{3z^{2}-r^{2}}$) are represented by t$_{2g}$ and e$_{g}$ band, respectively. In the VB, major contributions in t$_{2g}$ states are from the region -1.4 eV to -0.2 eV. In the region -8 eV to -3 eV very small intensity peaks of t$_{2g}$ and e$_{g}$ states are also observed. From PDOS of O atom (p orbital) shown in Fig. 2(c), it is observed that there are equal and opposite contributions in density of states from both the spin-up and spin-dn channel. It is also observed that there are negligible density of states from the Oxygen p orbital about the chemical potential ($\mu$ = 0 eV). The energy gap is created between d bands of V atom, which suggest that ZnV$_{2}$O$_{4}$ is a Mott-insulator.\cite{Imada}\\
   At the finite temperature, there is non-vanishing probability that some electrons will be thermally excited (across the energy gap E$_{g}$) to the lowest unoccupied conduction bands. The fraction of electrons excited across the gap at temperature T is roughly given by the order of e$^{-E_{g}/2\textit{k}_{B}T}$.\cite{Ashcroft} The fraction of thermally excited electrons in V atom, from region -8 eV to -3 eV, are $\sim$10$^{-44}$ at 300 K. Therefore, there are essentially negligible electrons excited across the gap from this region, and no contributions in the thermopower is expected. The intensity of t$_{2g}$ states are greater than the e$_{g}$ states in the entire range of energy shown in the Fig. 2(b). It is evident from the Fig. 2(b) that t$_{2g}$ states are at the top of the VB, whereas e$_{g}$ states are 175 meV lower than that of t$_{2g}$ states. At 300 K, the fraction of thermally excited electrons from t$_{2g}$ states (top of VB) to the unoccupied states of the CB are $\sim$5$\times$10$^{-4}$. For the e$_{g}$ states this factor is $\sim$1.6$\times$10$^{-5}$. The fraction of these thermally excited electrons from t$_{2g}$ states are almost 30 times greater than that of e$_{g}$. When the temperature will be 600 K, this fraction factor for t$_{2g}$ and e$_{g}$ states will be nearly 2.2$\times$10$^{-2}$ and 4$\times$10$^{-3}$, respectively. The temperature corresponding to the gap of 175 meV between t$_{2g}$ and e$_{g}$ band is $\sim$2100 K. At 2100 K, the thermally excited electrons from t$_{2g}$ and e$_{g}$ states of VB to the CB will be nearly equal. Therefore, in the temperature range 300-600 K, large number of electrons from t$_{2g}$ states will make transitions to the available unoccupied states at the bottom of the conduction band t$_{2g}$(e$_{g}$) states. Thus, the major contributions in thermopower of this compound are expected from the t$_{2g}$ states of the VB.\\ 
   The Experimentally observed value of Seebeck coefficient for ZnV$_{2}$O$_{4}$ in the temperature range 300-600 K is shown in Fig. 3.
    \begin{figure}[htbp]
  \begin{center}
    \includegraphics[clip,trim=0.0cm 0cm 0.0cm 0.0cm, width=0.42\textwidth]{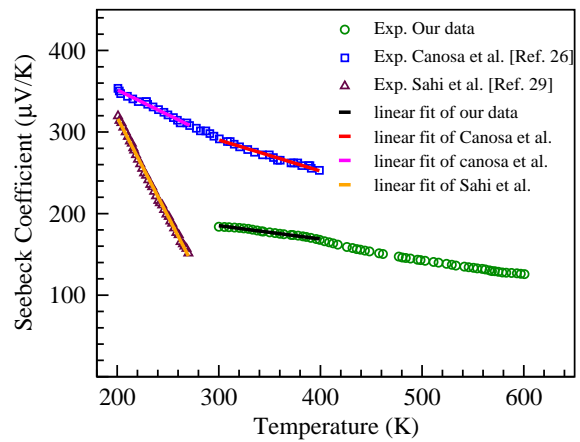}
    \label{}
    \captionsetup{justification=raggedright,
singlelinecheck=false
}
    \caption{(Color online) Temperature variations of Seebeck coefficient ($\alpha$) for ZnV$_{2}$O$_{4}$ compound.}
    \vspace{-0.3cm}
  \end{center}
\end{figure} 
   It is evident from the Fig. 3, the temperature dependent variation of $\alpha$ is almost linear. This behavior of $\alpha$ is similar to those reported in the literature.\cite{Canosa} The value of $\alpha$ is large and positive in the temperature range under study.   The positive $\alpha$ value indicates that holes are dominant charge carriers in the thermopower contribution for ZnV$_{2}$O$_{4}$ compound. The observed value of $\alpha$ at 300 K is $\sim$184 $\mu$V/K and its magnitude decreases continuously with increase in the averaged sample temperature.  At 600 K, the value of $\alpha$ is found to be $\sim$126 $\mu$V/K and a change of $\sim$58 $\mu$V/K is observed in its value in the temperature range 300-600 K. For comparison purpose, we have also inserted the reported data obtained by Canosa et al., in the Fig. 3.\cite{Canosa} At 300 K, the value of $\alpha$ obtained by Canosa et. al., is $\sim$292 $\mu$V/K, which is $\sim$108 $\mu$V/K larger in comparison to data obtained from our sample. For the temperature $\sim$400 K, we observed the $\alpha$ value $\sim$168 $\mu$V/K, which is $\sim$88 $\mu$V/K smaller than the value obtained by Canosa et. al., ($\sim$255 $\mu$V/K). In order to see the change in the $\alpha$ value with respect to temperature, we have done the linear fit in temperature range 300-400 K. In the temperature range 300-400 K, magnitudes of slope (d$\alpha$/dT) obtained from the fitting are $\sim$0.16 and $\sim$0.36 $\mu$V/K$^{2}$ for our sample and Canosa et al., data, respectively. The magnitude of d$\alpha$/dT for our sample in the temperature range 300-600 K (linear fitting line not shown here) is found to $\sim$0.21 $\mu$V/K$^{2}$. This clearly shows that for our sample the variation in $\alpha$ value with respect to temperature is smaller in comparison to the Canosa et. al., data. As per best of our knowledge, temperature dependent thermopower data is not reported for this compound above 400 K. In our sample, we have obtained the linear temperature dependent behavior upto 600 K temperature. In general the physical properties of oxide samples are very sensitive on sample synthesis conditions. The large difference in the magnitude of $\alpha$ is expected in oxide thermoelectric materials, especially in ZnV$_{2}$O$_{4}$, synthesized in different conditions, as the physical properties are highly sensitive to the sample preparation conditions.\cite{Reehuis, Zheng} In case of these materials (synthesized at high temperature), such a large difference may be due to the oxygen off-stoichiometry.\cite{Minh} Therefore, the observed difference in magnitude of $\alpha$ obtained from our sample and Canosa et. al., may be due to different oxygen off-stoichiometry between these two samples. A similar temperature dependent behaviour of $\alpha$ is also observed by Sahi et. al., in the temperature range 200-270 K.\cite{Shahi} The thermopower data obtained on ZnV$_{2}$O$_{4}$ compound by Sahi \textit{et al}, is shown in Fig. 3 and it is found that d$\alpha$/dT (magnitude of slope) is almost 4 times larger than that of Canosa \textit{et al}, in the temperature range 200-270 K. The magnitudes of $\alpha$ observed by Sahi \textit{et al}, are also found $\sim$34 and $\sim$162 $\mu$V/K smaller at temperature $\sim$200 and $\sim$270 K, respectively.\\
       In order to make comparison between our sample and Sahi et al. sample, we have estimated the structural parameters from X-ray diffraction patterns. The Rietveld analysis of room temperature ($\sim$300 K) X-ray diffraction data corresponding to space group Fd$\bar3$m gives lattice parameter a = 8.4086(5) $\AA$. The Wyckoff positions of Zn(0.125, 0.125, 0.125) and V(0.5, 0.5, 0.5) atom are fixed, whereas the refined positions of O atom obtained from the Rietveld analysis are (0.25890, 0.2589, 0.2589). For the same space group (Fd$\bar3$m) Reehuis et al. have also reported the lattice parameter, a = 8.4028(4) $\AA$, and the Wyckoff positions of O atom (0.2604, 0.2604, 0.2604) at 60 K. The lattice parameter obtained (corresponding to Fd$\bar3$m space group) by Sahi et al. is 8.4130(1)$\AA$ at 300 K, but they have not provided the refined Wyckoff positions of O atom.  The value of lattice parameter obtained for our compound is nearly 0.006 $\AA$ larger than that of Reehuis et al., whereas it is nearly 0.004 $\AA$ smaller than that of Sahi et al. The lattice parameter and refined Wyckoff positions are not mentioned by Canosa et al. Thus, it is difficult to make comparison of structural parameters of our sample with Canosa et al. The structural details reported for our sample can be a guideline for the future study of thermopower behavior of this compound. At this point, it is important to notice that the literature data used for comparison purpose is taken by using the data digitization technique. The digitization technique itself have an inherent error in the magnitude of a data extracted from the literature figure, therefore at this scale a maximum of 5 $\mu$V/K error is expected in the magnitude of a data point due to digitization. However, in the same temperature range i.e. 200-270 K, there is significantly large differences found between the Canosa et. al., and Sahi et. al., data  for the same compound. This suggest that sample synthesized in two different conditions can highly affect the thermopower value. It also gives a new direction of tuning the thermopower value in spinel oxides by sample preparation conditions.\\
 Furthermore, to analyze the experimental data and understand the large and positive Seebeck coefficient observed in ZnV$_{2}$O$_{4}$ compound, we have calculated the electronic band structure. The effective mass plays an important role to decide the sign and magnitude of Seebeck coefficient and it can be estimated from the electronic band structure. Under free electron theory approximations the relation between Seebeck coefficient ($\alpha$) and effective mass (\textit{m$^{*}$}) is given by the following formula.\cite{Snyder, Ashcroft}
  \begin{equation}
    \alpha = (8\varPi^{2} \text{\textit{k}}^2_B/3\textit{eh}^{2}) \textit{m}^{*}\textit{T}(\varPi/3\textit{n})^{2/3}
  \end{equation}
  where,  \textit{k$_{B}$}  is the Boltzmann constant and \textit{e} is the electronic charge, \textit{m$^{*}$} and \textit{n} are the effective mass of charge carrier and carrier density, respectively. The values of $\alpha$ and its sign are dependent on \textit{m$^{*}$}, temperature (T) and carriers density (\textit{n}). In the above expression of $\alpha$, T and \textit{n} are positive quantity, therefore the sign of $\alpha$ will be decided by the sign of \textit{m$^{*}$}, which depends on the band structure of materials. For semiconductor and insulator, the contribution in the $\alpha$ are from both type of charge carriers (i.e. electrons and holes). In the intrinsic semiconductor material, the value of carrier densities of \textit{n} and \textit{p-type} charge carriers are equal at a fixed temperature. So within the constant relaxation time approximation the sign of $\alpha$ will be decided by those charge carriers which have dominating effective mass. Thus, it is clearly evident from the Eq$^{n}$ (3) that the estimated value of effective mass from the dispersion curves can be helpful to make a qualitative understanding of the sign of experimentally observed value of $\alpha$.
   
   \begin{figure}[htbp]
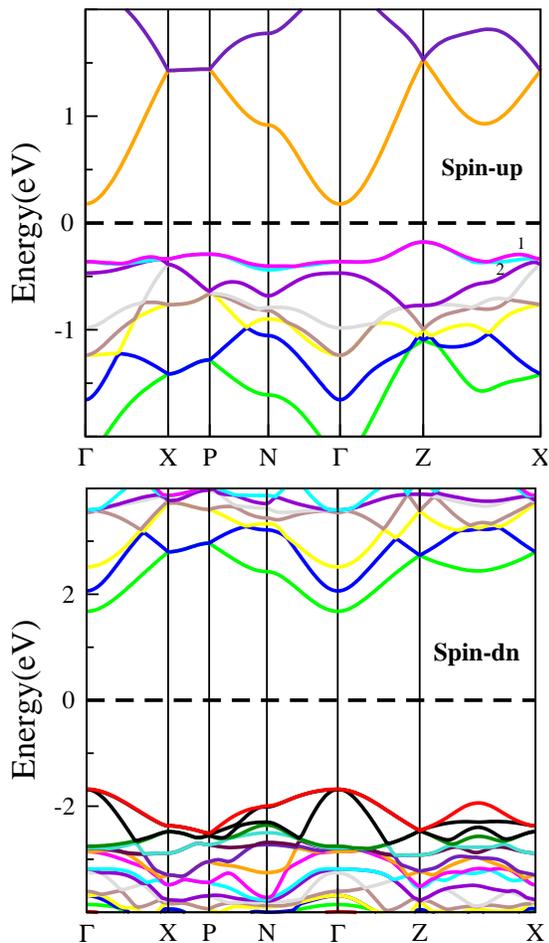

    \captionsetup[subfigure]{labelformat=empty}
\begin{center}

\subfloat[]{
        
        \includegraphics[clip,trim=0.cm 0cm 0.1cm 1.3cm, width=0.40\textwidth]{spinup.eps} } 
\vspace{-0.25cm}
\subfloat[]{
       
        \includegraphics[clip,trim=0cm 0.0cm 0cm 1.3cm, width=0.40\textwidth]{spindn.eps} } 
\captionsetup{justification=raggedright,
singlelinecheck=false
}
\caption{(Color online) Electronic band structure of ZnV$_{2}$O$_{4}$ compound, shown spin-up channel (top) and spin-down channel (bottom).}
\label{}
  \vspace{-0.2cm}
\end{center}
\end{figure}
    The spin polarized dispersion curves for spin-up and spin-dn channel are shown in Fig. 4. The energies for different k-values are calculated along the high symmetry directions ($\Gamma$-X-P-N-$\Gamma$-Z-X) in the 1st Brillouin zone. For the spin-up channel, the top of the valence band (VB) and bottom of the conduction band (CB) lies at Z and $\Gamma$-point, respectively. In case of spin-dn channel, top of the valence band and bottom of the conduction band lies at the same $\Gamma$ point. The computed energy gap for spin-up channel is found equal to the experimental energy band gap, which corresponds to the temperature of $\sim$3850 K. For the spin-dn channel, the estimated energy gap is $\sim$3.36 eV. This energy gap is corresponding to the temperature $\sim$40380 K. This is very large temperature and the material will be evaporated at such high temperature. Therefore, there are no contributions of charge carriers from the spin-dn channel in transport property (thermopower) of Zn$V_{2}$O$_{4}$ compound. Thus, the spin-up channel charge carriers in energy bands closer to the chemical potentials ($\mu$ = 0 eV) will mainly contribute in thermopower. From the dispersion curve, it is evident that ZnV$_{2}$O$_{4}$ compound is semiconductor with indirect band gap characteristics.\\
 At Z point, top of the valence band are doubly degenerate and it is index by band \textbf{1} and \textbf{2}. In CB, the first energy band closest to the chemical potential is non-degenerate at $\Gamma$ point. The degeneracy in  band \textbf{1} and \textbf{2}  is retained along Z-$\Gamma$ directions, whereas on further moving from $\Gamma$ to N point, degeneracy is completely lifted. The degeneracy is also lifted along the Z-X direction. The energy gap between top of the VB and bottom of the CB increases as one moves along Z-$\Gamma$ and Z-X directions, therefore the charge carriers present in top few states of VB ( band \textbf{1} and \textbf{2})  lies at Z point will participate in the thermopower contribution. At finite temperature, significant number of electrons will thermally excite across the energy gap (E$_{g}$), which results in creation of holes in the VB states. Therefore, from the VB both charge carriers, holes and electrons, are responsible in thermopower contribution. To know the type of charge carrier which is dominating to the observed value of thermopower in present compound, it is very essential to estimate the effective mass of charge carriers in VB and CB.\cite{Sharma} The shape of energy curve at a given k-point decides the value of effective mass, and under the parabolic approximation it is defined as,\cite{Ashcroft}
   \begin{equation}
     m^{*} = \hslash^{2}/(d^{2}E/dk^{2})
   \end{equation}
     The expression of Eq$^{n}$ 4, implies that effective mass for a flat energy curve will be greater than the narrower energy curve at a given k-point. From the energy curvature of bands \textbf{1} and \textbf{2} in VB lies at Z point, and 1st CB at $\Gamma$ point, it is expected that the effective mass value of the charge carriers will be larger in VB than the CB. Also, the bands \textbf{1} and \textbf{2} in VB are doubly degenerate, therefore the overall effective mass of holes present in these band will be large. The large and positive $\alpha$ value observed for Zn$V_{2}$O$_{4}$ compound is mainly due to the large effective mass of holes present in bands \textbf{1} and \textbf{2}.\\
 In order to see the effective mass contributions of holes and electrons in different high symmetric directions, we have calculated it at Z point for holes (along X, $\Gamma$, P, and N) and at $\Gamma$ point (along Z, X, P, and N) for electrons in VB and CB, respectively. The calculated effective mass is given in table I.
 \vspace{0.2cm}
 \begin{table}[htbp]
\caption{The effective mass of holes (at Z point) and electrons (at $\Gamma$ point) along high symmetric directions.  }   
\begin{tabular}{p{0.0cm}p{1.8cm}p{1.4cm}p{1.3cm}p{1.3cm}p{0.5cm}}
\hline
\hline
&&\multicolumn{2}{c}{Effective mass (m$^{*}$/m$_{e}$)}\\
\cline{3-4}
\\
&High-Symmetry Point&\underline{Valence Band}&&\underline{Conduction Band}\\
&&{ {\it } {Band \textbf{1}}}&{Band \textbf{2}}&&\\
\hline
Z$-$Z$\Gamma$&&2.11(7)&2.11(7)&-&\\
Z$-$ZX&&2.85(1)&2.71(2)&-&\\
Z$-$ZP&&3.65(7)&3.64(8)&-&\\
Z$-$ZN&&3.16(1)&2.99(7)&-&\\
$\Gamma$$-$$\Gamma$Z&&-&-&0.92(1)&\\
$\Gamma$$-$$\Gamma$X&&-&-&0.92(6)&\\
$\Gamma$$-$$\Gamma$P&&-&-&0.92(8)&\\
$\Gamma$$-$$\Gamma$N&&-&-&0.92(8)&\\

\hline
\hline
\end{tabular}
\end{table}
  In the table I, Z$-$Z$\Gamma$ represents the effective mass calculated at Z point along Z to $\Gamma$ direction, and similar notations for others. It is observed that effective mass of holes are larger than the electron$^{,}$s effective mass. The estimated value of effective mass for holes are 3.65 \textit{m}$_{e}$, along the P direction, which is maximum in comparison to other symmetric directions ($\Gamma$, X and N). The effective mass of electrons at $\Gamma$ point are found almost equal in Z, X, P and N directions. At Z point and along the P direction, the value of holes effective mass are nearly four times than the effective mass of electrons at $\Gamma$ point. The effective mass of holes are nearly three times larger than the electrons, along X and N directions. In the $\Gamma$ direction, holes effective mass is nearly double to the effective mass of electrons along the Z directions. The electrons moving from Z to P direction will create the holes in VB with maximum effective mass. It is also important to notice that at Z point, bands \textbf{1} and \textbf{2} are doubly degenerate. Therefore, the large contributions in Seebeck coefficient will be from effective mass of the holes. Thus, holes are the dominating charge carriers in thermopower contributions for Zn$V_{2}$O$_{4}$, and we observed the positive and large Seebeck coefficient for this compound. \\ 
From the electronic density of states and band structure studies, it is observed that Zn$V_{2}$O$_{4}$ compound have semiconducting ground state as well as large effective mass in doubly degenerate VB. These electronic properties give a direction to explore the thermoelectric properties of this compound, so that it can be used as thermoelectric material in high temperature applications. The calculated Seebeck coefficient for chemical potential ($\mu$) in the range -400 meV to 400 meV, at different temperatures are presented in Fig. 5. The positive and negative values of chemical potential represents the electron and hole doping, respectively. The $\alpha$ vs $\mu$ plots are at every 100 K temperature interval between 300-1400 K. The dashed line shown at $\mu$ = 0 meV, is fixed at middle of the gap.
\vspace{0.4cm}
\begin{figure}[htbp]
  \begin{center}
    \includegraphics[width=0.45\textwidth]{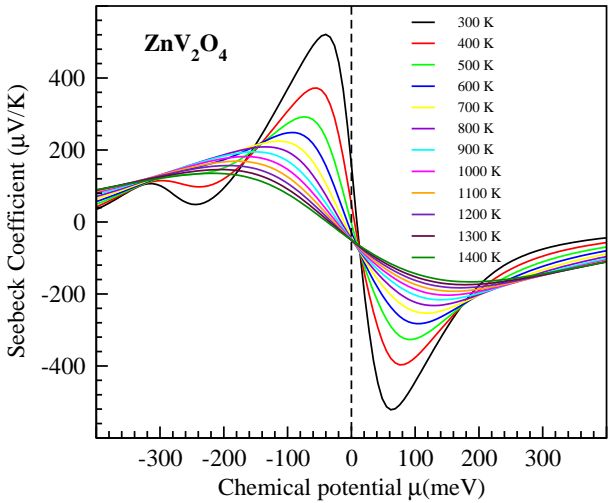}
    \label{}
    \captionsetup{justification=raggedright,
singlelinecheck=false
}
    \caption{(Color online) Variation of Seebeck coefficient ($\alpha$) with chemical potential ($\mu$) for ZnV$_{2}$O$_{4}$ compound at various temperature.}
    \vspace{-0.3cm}
  \end{center}
\end{figure} 
 For both positive and negative values of $\mu$, we observed a positive maximum value of $\alpha$ for a fixed temperature curve. At 300 K, the maximum values of $\alpha$ are $\sim$521 and -522 $\mu$V/K, for $\mu$ equal to -40 and 62 meV, respectively. As the temperature increases from 300 to 1400 K, the maximum value of $\alpha$ at a fixed temperature shifts away from the $\mu$ = 0 meV for the both positive and negative regions of the chemical potential. In the Fig. 5, chemical potential ($\mu$ = 0 meV), set at middle of the gap is corresponding to the ground state (T=0 K). From Eq$^{n}$ (2), it is clear that at a given finite temperature the value of chemical potential will shifted towards higher value by (3/4)\textit{k}$_{B}$T\textit{ln}(m$_{\textit{v}}$/m$_{\textit{c}}$). If we consider this finite temperature effect on chemical potential, the value of chemical potential will shift from $\mu$ = 0 meV by $\sim$20 meV at T = 300 K. At chemical potential shifted to $\sim$20 meV from $\mu$ = 0 meV , the calculated values of $\alpha$ still do not match with the experimentally observed data obtained at 300 K from our sample and Canosa et. al. However, including the temperature effect on chemical potential, at 300 K the experimentally obtained thermopower data from our sample and Canosa et. al, are positive and matches with the calculated values for the chemical potential $\sim$142 and $\sim$100 meV below the $\mu$= 0 meV, respectively. The matching of our sample data and Canosa et. al., data at two different chemical potential for a fixed temperature suggest that sample synthesized in two different conditions have different oxygen stoichiometry. From the chemical potential value it is evident that our sample have more oxygen off-stoichiometry in comparison to canosa et. al., or in other words our sample is more hole-doped than that of Canosa et. al., sample. Thus, the sample preparation condition can tune the oxygen off-stoichiometry in the oxide compound, and thus can tune the thermoelectric properties.
  From the above discussions, it is clear that the value of Seebeck coefficient is highly sensitive to the chemical potential. The value of chemical potential for a given oxide compound depends on the temperature, type of doping and the oxygen off-stoichiometry. In order to see the effect of these parameters on experimentally observed data, we have plotted the calculated and experimental data together. 
  \vspace{0.0cm}
  \begin{figure}[htbp]
  \begin{center}
    \includegraphics[width=0.42\textwidth]{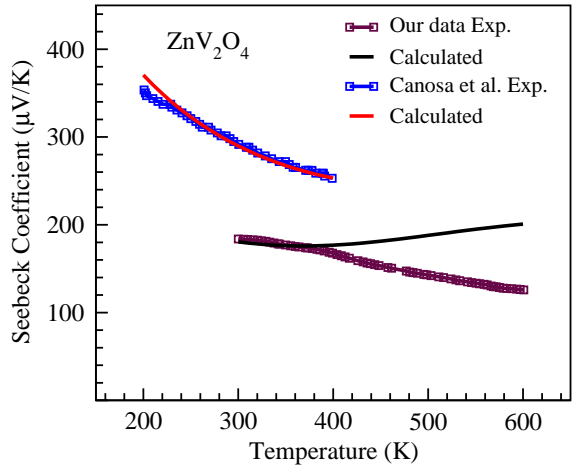}
    \label{}
    \captionsetup{justification=raggedright,
singlelinecheck=false
}
    \caption{(Color online) Temperature variation of thermopower ($\alpha$),  experimental ($\square$) and calculated ($-$), for ZnV$_{2}$O$_{4}$ compound.}
    \vspace{-0.1cm}
  \end{center}
\end{figure} 
 The temperature variations of thermopower is shown in the Fig. 6. The best possible matching between calculated and experimental data obtained by canosa et. al., and our data are found at $\mu$ equal to -125 and -162 meV, respectively. One can always tune the magnitude of $\alpha$ by adjusting the chemical potential.\\
 In the temperature range 230-400 K, calculated $\alpha$ values have reasonably good matching with the experimental data obtained by Canosa et. al. For the temperature below 230 K, calculated $\alpha$ values of present compound have a small deviation from experimental data. It has been earlier reported by Sharma et. al., that calculated transport coefficients below 200 K are very much sensitive to the k-point sampling of the Brillouin zone.\cite{Ssharma} Therefore, this deviation of calculated Seebeck coefficient values from Canosa et. al., data below 230 K can be attributed due to the insufficient k-point used in the present case. Currently, for anti-ferromagnetic structure having 28 atoms in the unit cell we have used the 27000 k-points in the full Brillouin zone. In the present case, the higher k-point sampling may give a better matching between experimental and calculated data below 230 K, but it will consume more time. For our sample we found a good matching between experimental and calculated data in the temperature range 300-410 K. Above 410 K, there is large difference between calculated and experimental data. The deviation of calculated data from the experimental may be due to the scattering effect at high temperature. Due to the various thermal scattering above the Debye temperature, relaxation time changes and this affect the calculated value of Seebeck coefficient in high temperature region.\\
 The selection of any material as a good thermoelectric material is also decided on the basis of the power factor (PF). The expression of \textit{ZT} (Eq$^{n}$ 1) implies that a material have higher \textit{ZT} if it has a large PF ($\alpha^{2}$$\sigma$). In order to see the possibilities of suitable doping and tuning the chemical potential to enhance the PF, we have also estimated the PF as a function of $\mu$ at various different temperature between 300-1400 K. At this point it is important to note that, in Fig 7. we have plotted the power factor with respect to scattering time, $\alpha^{2}$$\sigma$/$\tau$, as a function of chemical potential. If the experimental value of scattering time $\tau$ of the compound is known, then such plot will be helpful in the estimation of realistic value of PF. For the different temperature, the variation of $\alpha^{2}$$\sigma$/$\tau$ with chemical potential ($\mu$) ranging from -600 meV to 600 meV is shown in the Fig. 7. For every temperature there are two peaks with maximum $\alpha^{2}$$\sigma$/$\tau$ are found in the shown chemical potential range, one with positive $\mu$ and another with negative $\mu$. For the positive $\mu$, peak maximum value of the $\alpha^{2}$$\sigma$/$\tau$ at different temperature increases continuously with temperature from 300 to 1400 K. It is also found that the positions of peak maximum also gets shifted towards higher $\mu$ with increase in temperature. At 300 K,
\vspace{0.0cm}
\begin{figure}[htbp]
  \begin{center}
    \includegraphics[width=0.45\textwidth]{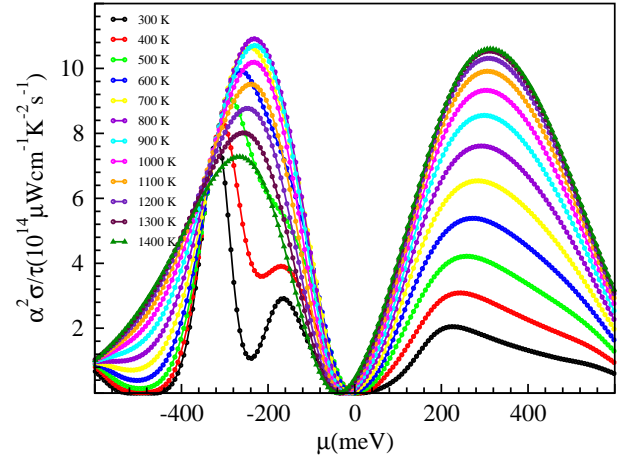}
    \label{}
    \captionsetup{justification=raggedright,
singlelinecheck=false
}
    \caption{(Color online) Variation of Power factor with respect to scattering time,$\alpha^{2}$$\sigma$/$\tau$, with chemical potential ($\mu$) for ZnV$_{2}$O$_{4}$ compound.}
    \vspace{-0.6cm}
  \end{center}
\end{figure} 
 the value of $\alpha^{2}$$\sigma$/$\tau$ is $\sim$2($\times$10$^{14}$ $\mu$Wcm$^{-1}$K$^{-2}$s$^{-1}$) for $\mu$ equal to $\sim$225 meV. The maximum value of $\alpha^{2}$$\sigma$/$\tau$ reaches to the value $\sim$10.60($\times$10$^{14}$ $\mu$Wcm$^{-1}$K$^{-2}$s$^{-1}$) at 1400 K, with higher value of $\mu$ equal to 312 meV. In the negative $\mu$ region, the maximum of $\alpha^{2}$$\sigma$/$\tau$ at 300 K is $\sim$7.80 ($\times$10$^{14}$ $\mu$Wcm$^{-1}$K$^{-2}$s$^{-1}$) for the $\mu$ value of $\sim$ -320 meV. For the temperature upto 800 K, the peak value at different temperature increases and in the region 800 to 1400 K, it decreases continuously with temperature. The value of $\alpha^{2}$$\sigma$/$\tau$ at peak maxima for 800 and 1400 K are equal to $\sim$10.90 and $\sim$7.30($\times$10$^{14}$ $\mu$Wcm$^{-1}$K$^{-2}$s$^{-1}$), respectively. In the negative region of chemical potential, the maxima of peak at different temperatures in the range 300-800 K gets shifted towards $\mu$ = 0 meV, the change in $\mu$ value is $\sim$89 meV. For the temperature range 800-1400 K, the peak position of maximum at different temperature has shifted towards the higher $\mu$ (in magnitude). In comparison to the positive region of the chemical potential, peak value of $\alpha^{2}$$\sigma$/$\tau$ at different temperature is found at lower $\mu$ in the negative range of the chemical potential. 
  From the Fig. 7, it is evident that the \textit{p-type} doping is more preferable to get the enhanced $\alpha^{2}$$\sigma$/$\tau$, whereas higher \textit{n-type} doping required to get the large value. We have also calculated the amount of charge necessary to move the position of chemical potential such that it can give the maximum $\alpha^{2}$$\sigma$/$\tau$, and hence the maximum PF. The estimated amount of charge carriers are found to be $\sim$0.12 electrons per formula unit of ZnV$_{2}$O$_{4}$. From the Fig. 3, it is clearly observed that around the top of VB, main contributions in the density of states comes from the V (\textit{3d}) electrons. Therefore, addition or removal of the charge carriers from the V atom is more suitable for tuning the thermoelectric properties. As discussed earlier, for better thermoelectric applications \textit{hole-type} doping is more appropriate than that of \textit{electron-type}. Therefore, it is suggested that \textit{hole-type} doping of an amount $\sim$0.06 charge per V atom can enhance the power factor in the doped ZnV$_{2}$O$_{4}$ compound. The experimentally observed value of ZnV$_{2}$O$_{4}$ compound is positive, therefore \textit{p-type} doping will enhance the PF of pure compound.\cite{Sonusharma}\\
   In order to check the potential capability of ZnV$_{2}$O$_{4}$ for thermoelectric applications, we have also calculated dimensionless parameter called as figure-of-merit (\textit{ZT}). The estimation of \textit{ZT} value for this compound gives an idea that how much efficiently this material can be used for thermoelectic applications. The \textit{ZT} value is calculated at different absolute temperature by using the Eq$^{n}$ (1). The temperature dependent variation of \textit{ZT} is shown in Fig. 8. 
\vspace{0.6cm}
  \begin{figure}[htbp]
  \begin{center}
    \includegraphics[width=0.4\textwidth]{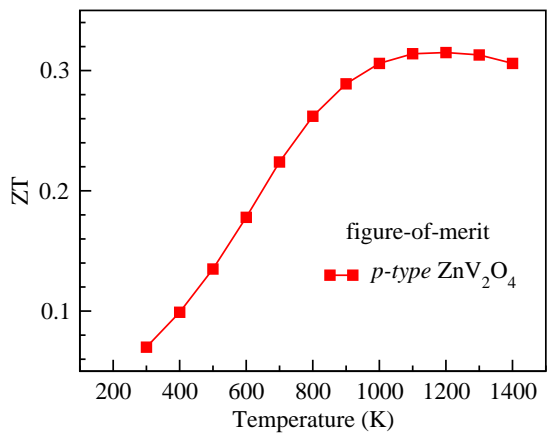}
    \label{}
    \captionsetup{justification=raggedright,
singlelinecheck=false
}
    \caption{(Color online) The dimensionless \textit{figure-of-merit} (\textit{ZT}) for \textit{p-type} ZnV$_{2}$O$_{4}$ compound.}
    \vspace{-0.5cm}
  \end{center}
\end{figure}  
 In the estimation of \textit{ZT}, we have used the value of power factor with respect to scattering time ($\alpha^{2}$$\sigma$/$\tau$) obtained from the BoltzTraP calculation. The value of $\tau$ is used as $\sim$10$^{-14}$ s. The thermal conductivity value ($\sim$3.33 W/mK) corresponding to 200 K temperature is taken as reported in the literature for the MnV$_{2}$O$_{4}$ and CoV$_{2}$O$_{4}$ compound.$^{19}$ At 300 K, the estimated value of \textit{ZT} is $\sim$0.07. As the temperature increases from 300 K to 900 K, the \textit{ZT} value increases almost linearly. At 900 K, \textit{ZT} is $\sim$0.29, on further increase in temperature it becomes almost constant up to 1400 K. From the Fig. 8, it is evident that in the temperature range 900-1400 K the value of \textit{ZT} ($\sim$ 0.3) is almost constant. The \textit{ ZT} value of this compound is less than half of the commercialized thermoelectric materials (Bi$_{2}$Te$_{3}$ and PbTe), though it is best suited to be used in the high temperature range due to its thermal stability. At this point it is important to notice that, the value of thermal conductivity, $\kappa$, used for computing the \textit{ZT} is taken from the single crystal materials. Further, the value of \textit{ZT} can be enhanced by decreasing the $\kappa$ value. It can be done by making the poly-crystalline sample in which the thermal conductivity is lowered due to scattering of thermal energy at grain boundaries.\cite{Tritt, Tmtritt} This way one can further increase the  \textit{ZT} value. The value of $\tau$ is taken constant, but at high temperature relaxation time decreases due to various scattering. Thus, this parameter also affect the \textit{ZT} value at higher temperature range. By taking the constant $\tau$ and available $\kappa$ value from literature, we have roughly estimated\textit{ZT} value. It shows that this material can be suitable for thermoelectric applications in the high temperature range.\\
 
\section{Conclusions} 
In the present work, we have studied the high temperature thermoelectric behaviour of ZnV$_{2}$O$_{4}$ compound. The experimental data of Seebeck coefficient is almost linear in the temperature range 300-600 K. The large and positive values of $\alpha$ observed at 300 and 600 K are found to be $\sim$184 and $\sim$126 $\mu$V/K, respectively. We have also studied the ground state electronic and thermoelectric properties of this compound by using the first principle calculations along with the BoltzTraP transport theory. The calculation on anti-ferromagnetic ground state structure of ZnV$_{2}$O$_{4}$ gives an energy gap equal to 0.33 eV, which is equal to the experimentally observed energy gap. From the energy band structure calculation, we have found that holes have large effective mass ($\sim$4 times),especially along the Z to P-direction, than that of electrons in conduction band. The large effective mass of holes in doubly degenerate bands (top of valence band) are mainly responsible to the large value of positive Seebeck coefficients observed in the compound. The calculated value of Seebeck coefficient is reasonably in good matching, in the temperature range 300-410 K, with experimentally observed data obtain from our sample. The experimental data reported in the literature also have good match in the large temperature range 230-400 K, for a particular chemical potential. The matching between calculated and experimental data in same temperature range but at two different chemical potential suggests that there is a possibility of tuning the thermopower value in the ZnV$_{2}$O$_{4}$ by controlling the synthesis conditions. The calculated power factor for this compound shows that the enhanced value of power factor can be found in this material with synthesis conditions as well as \textit{n} and \textit{p}-type doping. The calculation also suggest that \textit{p}-type doping is more preferable in comparison to \textit{n}-type. The estimated value of \textit{ZT} shows that ZnV$_{2}$O$_{4}$ compound can be used as good thermoelectric materials in high temperature range. The prediction of \textit{ZT} value is done by taking the constant value of thermal conductivity of single crystal spinel oxides. The value of thermal conductivity can be reduced by making the polycrystalline sample. Hence, further the \textit{ZT} value of the compound can be enhanced. Thus, our work shows that ZnV$_{2}$O$_{4}$ compound has a potential to be used as a good thermoelectric material in the high temperature range.

\end{document}